\documentclass[12pt]{article}
\usepackage{epsfig}
\usepackage{cite}
\usepackage{amsmath}
\usepackage{float}
\usepackage{wrapfig}
\usepackage{lipsum}
\usepackage{needspace}
\usepackage{ellipsis}
\usepackage{mathtools}

\newcommand{\mysection}{\setcounter{equation}{0}\section}

\def\beq{\begin{equation}}
\def\eeq{\end{equation}}
\def\beqa{\begin{eqnarray}}
\def\eeqa{\end{eqnarray}}
 
\newlength{\dinwidth} \newlength{\dinmargin}
\begin{document}

\begin{center}
{\Large \bf Resummation for $2\rightarrow n$ processes in single-particle-inclusive kinematics}
\end{center}
\vspace{2mm}
\begin{center}
{\large Matthew Forslund$^a$ and Nikolaos Kidonakis$^b$}\\
\vspace{2mm}
${}^a${\it Department of Physics and Astronomy, Stony Brook University, \\ 
Stony Brook, NY 11794, USA}

\vspace{1mm}

${}^b${\it Department of Physics, Kennesaw State University, \\
Kennesaw, GA 30144, USA}
\end{center}

\begin{abstract}
We present a formalism and detailed analytical results for soft-gluon resummation for $2\rightarrow n$ processes in single-particle-inclusive (1PI) kinematics. This generalizes previous work on resummation for $2 \rightarrow 2$ processes in 1PI kinematics. We also present soft anomalous dimensions at one and two loops for certain $2 \rightarrow 3$ processes involving top quarks and Higgs or $Z$ bosons, and we provide some brief numerical results.
\end{abstract}

\mysection{Introduction}

In theoretical calculations of hard-scattering cross sections of relevance to hadron colliders, the state of the art has been moving steadily towards higher orders, more loops, and resummations at higher logarithmic accuracy; it has also been gradually expanded to processes with larger numbers of final-state particles. In particular, soft-gluon resummations have become a very useful tool in making predictions for additional corrections beyond complete fixed-order results. The soft-gluon corrections appear in the perturbative series as logarithms of a threshold variable that involves the energy in the soft emission.

Soft-gluon resummation follows from factorization properties of the cross section \cite{NKGS,CLS,KOS,LOS,NKVD,BCMN} and it has been applied to a large number of processes in hadron collisions. Most of the applications for total cross sections and differential distributions have been done for $2 \rightarrow 2$ processes in single-particle-inclusive (1PI) as well as pair-invariant-mass (PIM) kinematics, most notably for top-quark production (see Ref. \cite{NKtoprev} for a review) but also many other processes. The choice of threshold variable in the resummation depends on the kinematics. For example, in PIM kinematics for top-antitop pair production, the threshold variable involves the invariant mass of the $t{\bar t}$ pair.

Applications to $2 \rightarrow 3$ processes using extensions of the PIM formalism, e.g. three-particle-invariant-mass kinematics, have also been made \cite{LLL,KMST,BFPSY,BFOP,BFPY,BFOPS,KMST2,KMSST,BFFPPT}. These processes include $t{\bar t}W$ production \cite{LLL,BFOP,KMSST,BFFPPT}, $t{\bar t}H$ production \cite{KMST,BFPSY,BFPY,KMST2,BFFPPT}, and $t{\bar t}Z$ production \cite{BFOPS,KMSST,BFFPPT}. In these extensions of the PIM formalism, the threshold variable involves the invariant mass of the three-particle final state, e.g. $t{\bar t}H$. 

In this paper, we instead generalize resummation to processes with $n$ particles in the final state explicitly in 1PI kinematics. In addition to providing an alternative way of calculating total cross sections, this new formalism also allows the calculation of 1PI differential distributions (for example in transverse momentum or rapidity) that cannot be calculated with the other kinematics. We also give more details for $2 \rightarrow 3$ processes with top quarks and Higgs or $Z$ bosons in 1PI kinematics.

In many cases, and especially for top-quark production (see Ref. \cite{NKtoprev} for a review of results in 1PI kinematics), these soft-gluon corrections are large; in fact, they numerically dominate the complete corrections and can be thought of as very good approximations to complete results. In particular, for top-antitop pair production, the soft-gluon corrections provide excellent approximations at next-to-leading order (NLO) and next-to-next-to-leading order (NNLO), and are significant even at next-to-NNLO (N$^3$LO) \cite{NKn3lo}. Another important set of processes where soft-gluon terms provide excellent approximations and large corrections involve single-top production \cite{NKst}, top production in association with a charged Higgs boson \cite{NKtH}, and top production via anomalous couplings in association with a $Z$ boson \cite{NKtZ}, a photon \cite{MFNK}, or a $Z'$ boson \cite{MGNK}.

We begin in Section 2 with the development of the formalism, starting with elementary considerations and kinematics for $2\rightarrow 2$ processes, and then for $2 \rightarrow 3$ processes, before moving on to the generalization to $2 \rightarrow n$ processes and the derivation of the resummed cross section in the general case. We define a threshold variable $s_{th}$ which measures the extra energy in soft radiation and which vanishes at partonic threshold. Logarithms of this threshold variable appear in the perturbative expansion as plus distributions of the general form $[\ln^m(s_{th}/s)/s_{th}]_+$, with $m \le 2n-1$  at $n$th order. The exponentiation of these threshold logarithms is organized in the resummed cross section. We also provide results for the expansion of the resummed cross section to fixed order, in particular NLO and NNLO. In Section 3, we provide some kinematical details about the cross section calculation at the partonic and hadronic levels. In Section 4 we present results for the soft anomalous dimensions through two loops for $2\rightarrow 3$ processes involving a top quark and a Higgs or $Z$ boson, and a brief numerical application to $t$-channel $tqH$ production which shows the power and relevance of the formalism. We conclude in Section 5, and we include two appendices with details on kinematical integration variables for $2 \rightarrow n$ processes.

\mysection{Resummation for $2 \rightarrow n$ processes}

In this section we develop the formalism for resummation in 1PI kinematics with multi-particle final states. We begin with some simple considerations and definitions for $2 \rightarrow 2$ processes in the next subsection, and extend them to $2 \rightarrow 3$ processes in subsection 2.2 and to $2 \rightarrow n$ processes in subsection 2.3. The complete resummation formalism for $2 \rightarrow n$ processes is given in subsection 2.4. Fixed-order expansions of the resummed cross section are provided in subsection 2.5. 

\subsection{Kinematics and threshold for $2 \rightarrow 2$ processes}

We first consider processes that are $2 \rightarrow 2$ at lowest order, $p_a +p_b \rightarrow p_1 +p_2$ (e.g. $q{\bar q} \rightarrow t {\bar t}$). We define the usual kinematical variables $s=(p_a+p_b)^2$, $t=(p_a-p_1)^2$, and $u=(p_b-p_1)^2$. We also define the threshold variable $s_{th}=s+t+u-p_1^2-p_2^2$. Of course $p_1^2=m_1^2$ and $p_2^2=m_2^2$ where, depending on the process, the masses $m_1$ and $m_2$ can be zero or finite. As we approach partonic threshold, $s_{th} \rightarrow 0$ and there is vanishing energy for any additional radiation.

If we have an additional gluon with momentum $p_g$ being emitted in the final state, then by using momentum conservation, $p_a+p_b=p_1+p_2+p_g$, it is straightforward to show that the above definition of $s_{th}$ is equivalent to $s_{th}=(p_2+p_g)^2-p_2^2$. It is clear that $s_{th}$ goes to 0 as $p_g$ goes to 0 (soft gluon). The physical meaning is also more clear from this way of writing $s_{th}$: it is the invariant mass squared of the ``particle 2 + gluon'' system minus the invariant mass squared of particle 2, i.e. it describes the extra energy in the soft emission. Note that particle 1 is the observed particle in this single-particle-inclusive kinematics.

If the incoming partons $a$ and $b$ come from hadrons $A$ and $B$, then we also define the hadron-level variables $S=(p_A+p_B)^2$, $T=(p_A-p_1)^2$, $U=(p_B-p_1)^2$, and $S_{th}=S+T+U-p_1^2-p_2^2$. Assuming that $p_a=x_a p_A$ and $p_b=x_b p_B$, where $x_a$ and $x_b$ denote the fraction of the momentum carried by partons $a$ and $b$ in hadrons $A$ and $B$, respectively, then we have the relations $s=x_a x_b S$, $t=x_aT+(1-x_a)p_1^2$, and $u=x_bU+(1-x_b)p_1^2$.

Then, using the above relations and after some algebra, we find that 
\beq
\frac{S_{th}}{S}=\frac{s_{th}}{s}-(1-x_a)\frac{\left(u-p_2^2\right)}{s}-(1-x_b)\frac{\left(t-p_2^2\right)}{s}+(1-x_a)(1-x_b) \frac{\left(p_1^2-p_2^2\right)}{s} \, .
\eeq
The last term, involving $(1-x_a)(1-x_b)$, is higher order and can be ignored near threshold, as $x_a\rightarrow 1$ and $x_b \rightarrow 1$.

\subsection{Kinematics and threshold for $2 \rightarrow 3$ processes}

We next consider processes that are $2 \rightarrow 3$ at lowest order, $p_a +p_b \rightarrow p_1 +p_2+p_3$ (e.g. $bq \rightarrow tq'H$). We define the parton-level variables $s$, $t$, $u$, and the hadron-level variables $S$, $T$, $U$, as before. If we have an additional gluon with momentum $p_g$ in the final state, then momentum conservation is $p_a +p_b=p_1 +p_2 +p_3+p_g$.

We can define the threshold variable as $s_{th}=(p_2+p_3+p_g)^2-(p_2+p_3)^2$. This clearly gives the same physical meaning as extra energy from gluon emission and clearly vanishes as $p_g \rightarrow 0$.  One can also show after some work that this is equivalent to $s_{th}=s+t+u-p_1^2-(p_2+p_3)^2$.

We also define $S_{th}=S+T+U-p_1^2-(p_2+p_3)^2$, and find, after some algebra, the relation
\beq
\frac{S_{th}}{S}=\frac{s_{th}}{s}-(1-x_a)\frac{\left(u-(p_2+p_3)^2\right)}{s}-(1-x_b)\frac{\left(t-(p_2+p_3)^2\right)}{s}
+(1-x_a)(1-x_b) \frac{\left(p_1^2-(p_2+p_3)^2\right)}{s} \, .
\eeq
The last term, involving $(1-x_a)(1-x_b)$, can be ignored in the threshold limit, as $x_a\rightarrow 1$ and $x_b \rightarrow 1$. We see that our results here are a natural extension of the relations for $2 \rightarrow 2$ kinematics.

\subsection{Kinematics and threshold for $2 \rightarrow n$ processes}

These relations can be extended to an arbitrary number of particles: we consider processes that are $2 \rightarrow n$ at lowest order, $p_a+p_b \rightarrow p_1+p_2+\cdots+p_n$. Again, we define the parton-level variables $s$, $t$, $u$, and the hadron-level variables $S$, $T$, $U$, as before.
With an additional gluon with momentum $p_g$ in the final state, momentum conservation is $p_a+p_b=p_1+p_2+\cdots+p_n+p_g$.  

Then the threshold variable is
$s_{th}=(p_2+\cdots+p_n+p_g)^2-(p_2+\cdots+p_n)^2$ with the same physical meaning as before, and vanishing as $p_g \rightarrow 0$. Using the abbreviation $p_{2\cdots n}=p_2+\cdots+p_n$, we can rewrite the threshold variable as $s_{th}=(p_{2\cdots n}+p_g)^2-p^2_{2\cdots n}$. We can also show that this variable can
also be written as $s_{th}=s+t+u-p_1^2-p^2_{2\cdots n}$.

We also define $S_{th}=S+T+U-p_1^2-p^2_{2\cdots n}$, and find that 
\beq
\frac{S_{th}}{S}=\frac{s_{th}}{s}-(1-x_a)\frac{\left(u-p^2_{2\cdots n}\right)}{s}-(1-x_b)\frac{\left(t-p^2_{2\cdots n}\right)}{s}
+(1-x_a)(1-x_b) \frac{\left(p_1^2-p^2_{2\cdots n}\right)}{s} \, .
\label{sth}
\eeq
Again, the last term, involving $(1-x_a)(1-x_b)$, can be ignored as $x_a\rightarrow 1$ and $x_b \rightarrow 1$.

Finally, we note that one can appropriately redefine the above relations if, instead of particle 1, the observed particle is $n$ or any of the other particles.

\subsection{Resummation}

The factorized form of the differential cross section in proton-proton collisions in 1PI kinematics is 
\beq
E_1\frac{d\sigma^{AB \rightarrow 1\cdots n}}{d^3p_1}=\sum_{a,b} \; 
\int dx_a \, dx_b \,  \phi_{a/A}(x_a) \, \phi_{b/B}(x_b) \, 
E_1 \frac{d{\hat \sigma}^{ab \rightarrow 1\cdots n}(s_{th})}{d^3p_1} \, ,
\label{facphi}
\eeq
where $E_1$ is the energy of the observed particle 1,  $\phi_{a/A}$ ($\phi_{b/B}$) are parton distribution functions (pdf) for parton $a$ ($b$) in proton $A$ ($B$), and ${\hat \sigma}^{ab \rightarrow 1\cdots n}$ is the hard-scattering partonic cross section. For simplicity we do not explicitly show in the above equation the dependence on $\mu_F$ and $\mu_R$, the factorization and renormalization scales.

The resummation of soft-gluon corrections follows from the factorization of the cross section in integral transform space \cite{NKGS,LOS}. We define Laplace transforms (indicated by a tilde) of the partonic cross section as 
${\tilde{\hat\sigma}}(N)=\int_0^s (ds_{th}/s) \,  e^{-N s_{th}/s} \, {\hat\sigma}(s_{th})$, where $N$ is the transform variable, and note that logarithms of $s_{th}$ transform into logarithms of $N$, with the latter exponentiating. We also define transforms of the pdf as ${\tilde \phi}(N)=\int_0^1 e^{-N(1-x)} \phi(x) \, dx$. These definitions are motivated by the structure of Eq. (\ref{sth}).

We also consider the parton-parton cross section $E_1 \, d\sigma^{ab \rightarrow 1\cdots n}/d^3p_1$, of the same form as Eq. (\ref{facphi}) but with the incoming hadrons replaced by partons \cite{NKGS,CLS,LOS,KOS,NKVD}  
\beq
E_1\frac{d\sigma^{ab \rightarrow 1\cdots n}(S_{th})}{d^3p_1}=
\int dx_a \, dx_b \,  \phi_{a/a}(x_a) \, \phi_{b/b}(x_b) \, 
E_1 \frac{d{\hat \sigma}^{ab \rightarrow 1\cdots n}(s_{th})}{d^3p_1} \, ,
\label{facpaphi}
\eeq
and define its transform (again indicated by a tilde) as 
\beq
E_1\frac{d{\tilde \sigma}^{ab \rightarrow 1\cdots n}(N)}{d^3p_1}=\int_0^S 
\frac{dS_{th}}{S} \,  e^{-N S_{th}/S} \, E_1\frac{d\sigma^{a b \rightarrow 1\cdots n}(S_{th})}{d^3p_1} \, . 
\label{csmom}
\eeq 
Taking a transform of Eq. (\ref{facpaphi}), as defined in Eq. (\ref{csmom}) above,  and using Eq. (\ref{sth}) (ignoring the higher-order terms), we have  
\beqa
E_1 \frac{d{\tilde \sigma}^{ab \rightarrow 1\cdots n}(N)}{d^3p_1} &=& \int_0^1 dx_a e^{-N_a (1-x_a)} \phi_{a/a}(x_a) \int_0^1 dx_b e^{-N_b (1-x_b)} \phi_{b/b}(x_b)
\nonumber \\ && \times
\int_0^s \frac{ds_{th}}{s} e^{-N s_{th}/s} E_1 \frac{d{\hat \sigma}^{ab \rightarrow 1\cdots n}(s_{th})}{d^3p_1}
\nonumber \\ 
&=& {\tilde \phi}_{a/a}(N_a) \, {\tilde \phi}_{b/b}(N_b) \, E_1 \frac{d{\tilde{\hat \sigma}}^{ab \rightarrow 1\cdots n}(N)}{d^3p_1} \, ,
\label{fac}
\eeqa
where $N_a=N(p^2_{2\cdots n}-u)/s$ and $N_b=N(p^2_{2\cdots n}-t)/s$.

Next, we proceed with a refactorization of the cross section in terms of a new set of functions \cite{NKGS,CLS,LOS,KOS,NKVD}. We first rewrite Eq. (\ref{sth}) as
\beqa
\frac{S_{th}}{S}& = & -(1-x_a)\frac{\left(u-p^2_{2\cdots n}\right)}{s}-(1-x_b)\frac{\left(t-p^2_{2\cdots n}\right)}{s}+\frac{s_{th}}{s}
\nonumber \\ 
&=& -w_a \frac{(u-p^2_{2\cdots n})}{s}- w_b \frac{(t-p^2_{2\cdots n})}{s}+w_S +\sum_{i=1}^n  w_i
\label{ws}
\eeqa
where the $w$'s denote dimensionless weights. Note that $w_a \neq 1-x_a$ and $w_b \neq 1-x_b$ since they refer to different functions.

Then, a refactorized form of this cross section \cite{NKGS,LOS,NKVD} is 
\beqa
E_1\frac{d{\sigma}^{ab \rightarrow 1\cdots n}}{d^3p_1}&=&\int dw_a \, dw_b \left(\prod_{i=1}^n dw_i \right) dw_S \, \psi_{a/a}(w_a) \, \psi_{b/b}(w_b) \left(\prod_{i=1}^n J_i(w_i) \right) 
\nonumber \\ && \times 
{\rm tr} \left\{H^{ab\rightarrow 1\cdots n}\left(\alpha_s(\mu_R)\right) \, 
S^{ab \rightarrow 1\cdots n}\left(\frac{w_S \sqrt{s}}{\mu_F} \right)\right\}
\nonumber \\ && \times
\delta\left(\frac{S_{th}}{S}+w_a\frac{(u-p^2_{2\cdots n})}{s}
+w_b \frac{(t-p^2_{2\cdots n})}{s}-w_S -\sum_{i=1}^n  w_i\right) \, .
\label{refact}
\eeqa
The infrared-safe hard function $H^{ab\rightarrow 1\cdots n}$ describes contributions from the amplitude and from the complex conjugate of the amplitude. 
The soft function $S^{ab \rightarrow 1\cdots n}$ describes the emission of noncollinear soft gluons in the $2\rightarrow n$ process. 
Both the hard and the soft functions are process-dependent matrices in color space in the partonic 
scattering, and the trace of their product is explicit in the above result. 
The functions $\psi$ are distributions for incoming partons at fixed value of momentum, that describe the dynamics of collinear emission from those partons, and differ from the pdf $\phi$ \cite{NKGS,CLS,KOS,LOS,GS}. The $J_i$ denote functions that describe collinear emission from final-state colored particles.

Taking a transform of Eq. (\ref{refact}), of the form defined in Eq. (\ref{csmom}), and using Eq. (\ref{ws}), we then
have
\beqa
E_1\frac{d{\tilde \sigma}^{ab \rightarrow 1\cdots n}(N)}{d^3p_1}&=& 
\int_0^1 dw_a e^{-N_a w_a} \psi_{a/a}(w_a) \int_0^1 dw_b e^{-N_b w_b} \psi_{b/b}(w_b)
\nonumber \\ && \hspace{-25mm} \times
\left(\prod_{i=1}^n \int_0^1 dw_i e^{-N w_i} J_i(w_i) \right) 
{\rm tr}\left\{H^{ab \rightarrow 1\cdots n}\left(\alpha_s(\mu_R)\right) \int_0^1 dw_s e^{-N w_s}  
S^{ab \rightarrow 1\cdots n}\left(\frac{w_s\sqrt{s}}{\mu_F} \right)\right\} 
\nonumber \\ && \hspace{-25mm} 
={\tilde \psi}_{a/a}(N_a) \, {\tilde \psi}_{b/b}(N_b) \left(\prod_{i=1}^n  {\tilde J_i} \left(N \right)\right) 
{\rm tr} \left\{H^{ab \rightarrow 1\cdots n}\left(\alpha_s(\mu_R)\right) \, 
{\tilde S}^{ab \rightarrow 1\cdots n}\left(\frac{\sqrt{s}}{N \mu_F} \right)\right\} \, .
\nonumber \\
\label{refac}
\eeqa
We note that in this refactorized form all the $N$-dependence is absorbed in the functions ${\tilde S}$, ${\tilde \psi}$, and ${\tilde J}$, while the hard function $H$ is independent of $N$; this is in contrast to the original factorized form where both the partonic cross section ${\tilde {\hat \sigma}}$ and the parton densities ${\tilde \phi}$ are $N$-dependent \cite{NKGS,CLS,KOS,LOS,GS}.

Comparing Eqs. (\ref{fac}) and (\ref{refac}), we get the following expression for the transform-space hard-scattering partonic cross section,
\beq
E_1\frac{d{\tilde{\hat \sigma}}^{ab \rightarrow 1\cdots n}(N)}{d^3p_1}=
\frac{{\tilde \psi}_a(N_a) \, {\tilde \psi}_b(N_b) \, \left(\prod_{i=1}^n  {\tilde J_i} (N)\right)}{{\tilde \phi}_{a/a}(N_a) \, 
{\tilde \phi}_{b/b}(N_b)} {\rm tr} \left\{H^{ab\rightarrow 1\cdots n}\left(\alpha_s(\mu_R)\right) \, 
{\tilde S}^{ab \rightarrow 1\cdots n}\left(\frac{\sqrt{s}}{N \mu_F} \right)\right\} \, .
\label{sigN}
\eeq

The $N$-dependence of the soft matrix ${\tilde S}^{ab \rightarrow 1\cdots n}$ is resummed via renormalization group evolution\cite{NKGS}. We have 
\beq
{\tilde S}_b^{ab\rightarrow 1\cdots n}=(Z_S^{ab \rightarrow 1\cdots n})^{\dagger} \; {\tilde S}^{ab \rightarrow 1\cdots n} \; Z_S^{ab\rightarrow 1\cdots n}
\eeq
where ${\tilde S}_b^{ab\rightarrow 1\cdots n}$ is the unrenormalized quantity 
and $Z_S^{ab\rightarrow 1\cdots n }$ is a matrix of renormalization constants.
Thus, ${\tilde S}^{ab \rightarrow 1\cdots n}$ obeys the renormalization group equation
\beq
\left(\mu_R \frac{\partial}{\partial \mu_R}
+\beta(g_s)\frac{\partial}{\partial g_s}\right) {\tilde S}^{ab\rightarrow 1\cdots n}
=-(\Gamma_S^{ab\rightarrow 1\cdots n})^{\dagger} \; {\tilde S}^{ab\rightarrow 1\cdots n}-{\tilde S}^{ab\rightarrow 1\cdots n} \; 
\Gamma_S^{ab\rightarrow 1\cdots n}
\eeq
where $g_s^2=4\pi\alpha_s$ and $\beta$ is the QCD beta function,
\beq
\beta(\alpha_s)=\frac{d\ln\alpha_s}{d\ln\mu_R^2}=-\sum_{n=0}^{\infty}\beta_n \left(\frac{\alpha_s}{4\pi}\right)^{n+1} \, .
\eeq
The lowest-order term in the above series for the beta function  \cite{GW,HDP} is given by $\beta_0=(11C_A-2n_f)/3$ where $C_A=N_c$, with $N_c$ the number of colors, and $n_f$ is the number of light quark flavors. The evolution of the soft function is controlled by the soft anomalous dimension matrix, $\Gamma_S^{ab\rightarrow 1\cdots n}$, which is calculated from the coefficients of the ultraviolet poles of eikonal diagrams \cite{NKGS,KOS,NKst,NK2loop,NK3loop}.

The transform-space resummed cross section is derived from the renormalization-group evolution of the soft function and the other $N$-dependent functions in Eq. (\ref{sigN}), 
and it is given by \cite{NKGS,LOS,NKtoprev}
\beqa
E_1\frac{d{\tilde{\hat \sigma}}^{ab\rightarrow 1\cdots n}_{\rm resum}(N)}{d^3p_1} &=&
\exp\left[ \sum_{i=a,b} E_{i}(N_i)\right] \, 
\exp\left[ \sum_{i=a,b} 2 \int_{\mu_F}^{\sqrt{s}} \frac{d\mu}{\mu} \gamma_{i/i}(N_i)\right] \, 
\exp\left[ \sum_{i={\rm f.s.} \, q,g} \!\! E'_i(N)\right]
\nonumber\\ && \times \,
{\rm tr} \left\{H^{ab\rightarrow 1\cdots n}\left(\alpha_s(\sqrt{s})\right) {\bar P} \exp \left[\int_{\sqrt{s}}^{{\sqrt{s}}/N}
\frac{d\mu}{\mu} \; \Gamma_S^{\dagger \, ab\rightarrow 1\cdots n}\left(\alpha_s(\mu)\right)\right] \; \right.
\nonumber\\ && \left. \hspace{10mm} \times \,
{\tilde S}^{ab\rightarrow 1\cdots n} \left(\alpha_s\left(\frac{\sqrt{s}}{N}\right)\right) \;
P \exp \left[\int_{\sqrt{s}}^{{\sqrt{s}}/N}
\frac{d\mu}{\mu}\; \Gamma_S^{ab\rightarrow 1\cdots n}
\left(\alpha_s(\mu)\right)\right] \right\} 
\nonumber \\
\label{resummed}
\eeqa
where the symbols $P$ (${\bar P}$) refer to
path-ordering in the same (reverse) sense as the integration variable $\mu$.

The first exponential resums universal soft and collinear contributions from the incoming partons \cite{GS,CT}, 
\beq
E_i(N_i)=
\int^1_0 dz \frac{z^{N_i-1}-1}{1-z}\;
\left \{\int_1^{(1-z)^2} \frac{d\lambda}{\lambda}
A_i\left(\alpha_s(\lambda s)\right)
+D_i\left[\alpha_s((1-z)^2 s)\right]\right\} \, ,
\label{Eexp}
\eeq
with $A_i = \sum_{k=1}^{\infty} (\alpha_s/\pi)^k A_i^{(k)}$, where 
$A_i^{(1)}=C_i$ with $C_i=C_F=(N_c^2-1)/(2N_c)$ for a quark 
or antiquark and $C_i=C_A$ for a gluon, 
while $A_i^{(2)}=C_i K/2$ with $K= C_A\; ( 67/18-\pi^2/6 ) - 5n_f/9$.    
Also $D_i=\sum_{k=1}^{\infty} (\alpha_s/\pi)^k D_i^{(k)}$, 
with $D_i^{(1)}=0$ in Feynman gauge ($D_i^{(1)}=-A_i^{(1)}$ in axial gauge).
The second exponential gives the scale evolution in terms of the parton anomalous dimensions $\gamma_{i/i}=-A_i \ln N_i+\gamma_i$ where $\gamma_i=\sum_{k=1}^{\infty} (\alpha_s/\pi)^k \gamma_i^{(k)}$, with $\gamma_q^{(1)}=3C_F/4$ for quarks and 
$\gamma_g^{(1)}=\beta_0/4$ for gluons.

The exponential involving $E'_i$ describes radiation from any final-state (f.s.) massless quarks and gluons \cite{KOS,LOS}. The exponential is of course absent for colorless particles, and it is also absent for massive particles since the mass protects against mass divergences. For final-state massless quarks or gluons we have
\beq
E'_i(N)=
\int^1_0 dz \frac{z^{N-1}-1}{1-z}\;
\left \{\int^{1-z}_{(1-z)^2} \frac{d\lambda}{\lambda}
A_i \left(\alpha_s\left(\lambda s\right)\right)
+B_i\left[\alpha_s((1-z)s)\right]
+D_i\left[\alpha_s((1-z)^2 s)\right]\right\} \, ,
\label{E'exp}
\eeq
where $B_i=\sum_{k=1}^{\infty} (\alpha_s/\pi)^k B_i^{(k)}$, 
with $B_q^{(1)}=-3C_F/4$ for quarks and $B_g^{(1)}=-\beta_0/4$ for gluons.

We note that for jet production the final-state exponential can have different forms depending on definitions or constraints for the jets \cite{KOS}. In this paper we do not study jet or hadron production but focus on single-particle-inclusive cross sections, with the form of the exponent for the final-state particles as given in Eq. (\ref{E'exp}) above.

The process-dependent hard and soft functions (matrices) have the perturbative expansions 
$H^{ab\rightarrow 1\cdots n}=\sum_{k=0}^{\infty}(\alpha_s^{d+k}/\pi^k) H^{(k)}$, where the power $d$ depends on the partonic process, 
and ${\tilde S}^{ab\rightarrow 1\cdots n}=\sum_{k=0}^{\infty} (\alpha_s/\pi)^k {\tilde S}^{(k)}$. Finally the soft anomalous dimension has the expansion $\Gamma_S^{ab\rightarrow 1\cdots n}=\sum_{k=1}^{\infty}(\alpha_s/\pi)^k \Gamma_S^{(k)}$.

The moment-space resummed cross section in Eq. (\ref{resummed}) resums logarithms of the moment variable $N$. The logarithmic accuracy of the resummed cross section in Eq. (\ref{resummed}) is not a priori limited, but it depends on how many higher-order terms are included in the exponentials and in the process-dependent hard and soft functions and soft anomalous dimensions. When the (next-to-)leading powers of logarithms of $N$ are resummed, then we have (next-to-)leading-logarithm accuracy, etc. For next-to-leading-logarithm (NLL) resummation, we need one-loop results for the process-dependent functions; for next-to-NLL (NNLL) resummation, we need two-loop results, etc. When we invert the resummed cross section back to momentum space we get powers of logarithms of $s_{th}$ in the form of plus distributions, with the exact form given in the next subsection.

\subsection{Fixed-order expansions}

We can expand the formula for the resummed cross section, Eq. (\ref{resummed}), to any fixed order and invert it back to momentum space. Below we provide explicit results for the soft-gluon corrections at NLO and NNLO. 

The NLO soft-gluon corrections are
\beq
E_1\frac{d{\hat{\sigma}}^{(1)}}{d^3p_1} = F_{LO} \frac{\alpha_s(\mu_R)}{\pi}
\left\{c_3\, {\cal D}_1(s_{th}) + c_2\,  {\cal D}_0(s_{th}) 
+c_1\,  \delta(s_{th})\right\}+\frac{\alpha_s^{d+1}(\mu_R)}{\pi} 
\left[A^c \, {\cal D}_0(s_{th})+T_1^c \, \delta(s_{th})\right] \, ,
\label{NLOmaster}
\eeq
where the plus distributions of logarithms of the threshold variable are denoted by
\beq
{\cal D}_k(s_{th})=\left[\frac{\ln^k(s_{th}/s)}{s_{th}}\right]_+ \, . 
\eeq
Here $F_{LO}=\alpha_s^d \, {\rm tr}\{H^{(0)} S^{(0)}\}$ denotes the leading-order (LO) coefficient, 
\beq
c_3=2 (A_a^{(1)}+A_b^{(1)}) -\sum_{i={\rm f.s.} \, q,g} \!\!\! A_i^{(1)}\, ,
\label{c3}
\eeq 
where the sum in the last term is over final-state massless quarks and gluons, 
and $c_2$ is given by $c_2=c_2^{\mu}+T_2$, 
with
\beq
c_2^{\mu}=-(A_a^{(1)}+A_b^{(1)}) \ln\left(\frac{\mu_F^2}{s}\right)
\eeq
denoting the terms involving logarithms of the scale, and  
\beq
T_2=-2 \, A_a^{(1)} \, \ln\left(\frac{-u+p^2_{2\cdots n}}{s}\right)
-2 \, A_b^{(1)} \, \ln\left(\frac{-t+p^2_{2\cdots n}}{s}\right) +D_a^{(1)}+D_b^{(1)}
+\sum_{i={\rm f.s.} \, q,g} \!\! \left(B_i^{(1)}+D_i^{(1)}\right) \, 
\label{T2}
\eeq
denoting the scale-independent terms.
Also,
\beq
A^c={\rm tr} \left(H^{(0)} \Gamma_S^{(1)\,\dagger} S^{(0)}
+H^{(0)} S^{(0)} \Gamma_S^{(1)}\right) \, .
\label{Ac}
\eeq

With regard to the $\delta(s_4)$ terms, we split them into a term
$c_1$, that is proportional 
to the Born cross section, and a term $T_1^c$ that is not.
We write $c_1 =c_1^{\mu} +T_1$, with
\beq
c_1^{\mu}=\left[A_a^{(1)}\, \ln\left(\frac{-u+p^2_{2\cdots n}}{s}\right) 
+A_b^{(1)}\, \ln\left(\frac{-t+p^2_{2\cdots n}}{s}\right)
-\gamma_a^{(1)}-\gamma_b^{(1)}\right]\ln\left(\frac{\mu_F^2}{s}\right)
+d \frac{\beta_0}{4} \ln\left(\frac{\mu_R^2}{s}\right) 
\label{c1mu}
\eeq
denoting the terms involving logarithms of the scale.
We note that $T_1$ and $T_1^c$ cannot be calculated from the resummation formalism but they can be determined from a comparison to a complete NLO calculation.

We also note that these results differ from past expressions for $2 \rightarrow 2$ processes (see e.g. the review in Ref. \cite{NKtoprev} or the earlier review, using different notation, in Ref. \cite{NKmpla}) by having a generalized argument of the logarithms involving $u$ and $t$ in Eqs. (\ref{T2}) and (\ref{c1mu}), and by having an expanded sum over final-state particles in Eqs. (\ref{c3}) and (\ref{T2}). Of course, the $2\rightarrow n$ expressions reduce to the $2\rightarrow 2$ expressions when $n=2$.

The NNLO soft-gluon corrections are 
\beqa
E_1\frac{d{\hat{\sigma}}^{(2)}}{d^3p_1}&=&F_{LO} \frac{\alpha_s^2(\mu_R)}{\pi^2}
\left\{\frac{1}{2}c_3^2\, {\cal D}_3(s_{th}) + 
\left[\frac{3}{2}c_3 c_2-\frac{\beta_0}{4} c_3
+\frac{\beta_0}{8} \sum_{i={\rm f.s.} \, q,g} \!\!\! A_i^{(1)}\right]  {\cal D}_2(s_{th}) \right.
\nonumber \\ && \hspace{-21mm}
{}+\left[c_3 c_1+c_2^2-\zeta_2 c_3^2
-\frac{\beta_0}{2} T_2+\frac{\beta_0}{4} c_3 
\ln\left(\frac{\mu_R^2}{s}\right)+2(A_a^{(2)}+A_b^{(2)})
+\sum_{i={\rm f.s.} \, q,g} \!\! \left(-A_i^{(2)}+\frac{\beta_0}{4} B_i^{(1)}\right)\right] {\cal D}_1(s_{th})
\nonumber \\ && \hspace{-21mm} 
{}+\left[c_2 c_1-\zeta_2 c_3 c_2+\zeta_3 c_3^2
+\frac{\beta_0}{4} c_2 \ln\left(\frac{\mu_R^2}{s}\right) 
-\frac{\beta_0}{2} A_a^{(1)} \ln^2\left(\frac{-u+p^2_{2\cdots n}}{s}\right)
-\frac{\beta_0}{2} A_b^{(1)} \ln^2\left(\frac{-t+p^2_{2\cdots n}}{s}\right)
\right. 
\nonumber \\ && \hspace{-13mm} 
{}+\left(-2 A_a^{(2)}+\frac{\beta_0}{2} D_a^{(1)}\right) 
\ln\left(\frac{-u+p^2_{2\cdots n}}{s}\right)
+\left(-2 A_b^{(2)}+\frac{\beta_0}{2} D_b^{(1)}\right) 
\ln\left(\frac{-t+p^2_{2\cdots n}}{s}\right)
\nonumber \\ && \hspace{-13mm}  
{}+D_a^{(2)}+D_b^{(2)}+\frac{\beta_0}{8} (A_a^{(1)}+A_b^{(1)}) \ln^2\left(\frac{\mu_F^2}{s}\right)-(A_a^{(2)}+A_b^{(2)}) \ln\left(\frac{\mu_F^2}{s}\right)
\nonumber \\ && \hspace{-13mm}  \left. \left. 
{}+\sum_{i={\rm f.s.} \, q,g} \!\! \left(B_i^{(2)}+D_i^{(2)}\right) \right] 
 {\cal D}_0(s_{th}) \right\}
\nonumber \\ && \hspace{-23mm}
{}+\frac{\alpha_s^{d+2}(\mu_R)}{\pi^2} 
\left\{\frac{3}{2} c_3 A^c \, {\cal D}_2(s_{th})
+\left[\left(2 c_2-\frac{\beta_0}{2}\right) A^c+c_3 T_1^c +F^c\right]
{\cal D}_1(s_{th}) \right. 
\nonumber \\ && \hspace{3mm} \left. 
{}+\left[\left(c_1-\zeta_2 c_3+\frac{\beta_0}{4}\ln\left(\frac{\mu_R^2}{s}
\right)\right)A^c+c_2 T_1^c  +G^c\right]
{\cal D}_0(s_{th}) \right\} \, ,
\label{NNLOmaster}
\eeqa
where
\beq
F^c={\rm tr} \left[H^{(0)} \left(\Gamma_S^{(1)\,\dagger}\right)^2 S^{(0)}
+H^{(0)} S^{(0)} \left(\Gamma_S^{(1)}\right)^2
+2 H^{(0)} \Gamma_S^{(1)\,\dagger} S^{(0)} \Gamma_S^{(1)} \right] 
\label{Fterm}
\eeq
and
\beq
G^c={\rm tr} \left[H^{(1)} \Gamma_S^{(1)\,\dagger} S^{(0)}
+H^{(1)} S^{(0)} \Gamma_S^{(1)} + H^{(0)} \Gamma_S^{(1)\,\dagger} S^{(1)}
+H^{(0)} S^{(1)} \Gamma_S^{(1)} +H^{(0)} \Gamma_S^{(2)\,\dagger} S^{(0)}
+H^{(0)} S^{(0)} \Gamma_S^{(2)} \right] \, .
\eeq
Again, these results generalize expressions for $2\rightarrow 2$ processes (see e.g. \cite{NKtoprev}) and reduce to them when $n=2$.
We note that at NNLL (or higher) resummation accuracy for a given process, all soft-gluon terms in the expansion through NNLO can be fully calculated. 

\mysection{Cross section and kinematics}

In this section we provide some formulas that are needed for the calculation of cross sections with multi-particle final states.

It has been shown by Byckling and Kajantie \cite{BKPartKinematics,BKRed} that one can write the expression for the phase space integration of a $2\rightarrow n$ scattering process while integrating over only invariant variables. The details are given in Appendix A. One can alternatively \cite{BKRed} do the phase space integration by breaking the process down into successive $1\rightarrow 2$ decays and integrating over the relevant solid angle in each rest frame explicitly, as shown in Appendix B. Either way one obtains an expression for the differential partonic cross section $d^2\hat{\sigma}^{ab\rightarrow 1\cdots n}/(dt_{n-1}du_{n-1})$ where $t_{n-1} = (p_a-p_1-\cdots-p_{n-1})^2$ and $ u_{n-1} = (p_b-p_1-\cdots-p_{n-1})^2$.

The LO hadronic cross section is obtained by convoluting the differential partonic cross section with the appropriate parton distribution functions:
\begin{equation}
S^2\frac{d^2\sigma^{pp\rightarrow 1\cdots n}}{dT_{n-1}dU_{n-1}} = \int_{x_a^-}^1\frac{dx_a}{x_a}\int_{x_b^-}^1\frac{dx_b}{x_b}\phi(x_a)\phi(x_b)s^2\frac{d^2\hat{\sigma}^{ab\rightarrow 1\cdots n}}{dt_{n-1}du_{n-1}} \, ,
\end{equation}
where $S$, $T_{n-1}$, and $U_{n-1}$ are the hadronic analogues of the partonic invariants. We extend $2\rightarrow 3$ particle kinematic definitions \cite{Beenakker:1990maa} to $2\rightarrow n$ particle kinematics, giving the conditions
\begin{align}
	&t_{n-1}=x_b(T_{n-1}-m_n^2)+m_n^2 \, , & &u_{n-1}=x_a(U_{n-1}-m_n^2)+m_n^2 \, , & &s=x_ax_bS \, , \nonumber \\
    &s+t_{n-1}+u_{n-1}-m_n^2\ge\sum_{i=1}^{n-1}m_i^2 \, , & & 0\le x_a,x_b\le1 \, ,
\end{align}
which yield the integration bounds for $x_a$ and $x_b$:{}
\begin{align}
	&x_{a}^- = \frac{-T_{n-1}+\sum_{i=1}^{n-1} m_i^2}{S+U_{n-1}-m_n^2} \, , & &x_b^-=\frac{-m_n^2-x_a (U_{n-1}-m_n^2)+\sum_{i=1}^{n-1}m_i^2}{x_aS+T_{n-1}-m_n^2} \, .
\end{align}

For an arbitrary $2\rightarrow n$ process, there are $\frac{1}{2} (n-2)(n-3)$ relations between all possible kinematic invariants that are not fixed by momentum conservation. These must instead be fixed by the condition that any five or more vectors are always linearly dependent in four-dimensional space and their symmetric Gram determinant vanishes:
\begin{equation}
	\Delta_{l+1}(p_1,p_2,\cdots,p_l,-p_b)=0 \, , \qquad 4\leq l \leq n \, .
\end{equation}

The Gram determinant condition $\Delta_{l+1}=0$ can be equivalently written as a Cayley determinant condition \cite{BKPartKinematics} as
\begin{equation}
 \Delta_{l+1}(p_1,p_2,\cdots,p_l,-p_b) = \begin{vmatrix}
 0 & 1 & 1 & 1 & \cdots & 1 & 1 \\
 1 & 0 & p_1^2 & p_{12}^2 & \cdots & p_{12\ldots l}^2 & 0 \\
 1 & p_1^2 & 0 & p_{2}^2 & \cdots & p_{23\ldots l}^2 &  t_1\\
 1 &  p_{12}^2 & p_2^2 & 0 & \cdots & p_{34\ldots l}^2 & t_2\\
 \vdots & \vdots & \vdots & \vdots &  & \vdots & \vdots \\
 1 & p_{12\ldots l}^2 & p_{23\ldots l}^2 & p^2_{34\ldots l} & \cdots & 0 & t_{l} \\
 1 & 0 & t_1 & t_2 & \cdots & t_{l} & 0
\end{vmatrix} = 0 \, .
\end{equation}

\mysection{Soft-gluon corrections for $2 \rightarrow 3$ processes with a top quark and a Higgs or $Z$ boson}

In this section we consider several processes involving a three-particle final state with a top quark and a Higgs boson, or a top quark and a $Z$ boson.
We present the soft anomalous dimension matrices for these processes at one and two loops.  We also give some brief numerical results for $t$-channel $tqH$ production to illustrate the use of the formalism. 

We begin with the $s$-channel processes
$q(p_a)+{\bar q'}(p_b) \rightarrow t(p_1) +{\bar b}(p_2)+H(p_3)$
and $q(p_a)+{\bar q'}(p_b) \rightarrow t(p_1) +{\bar b}(p_2)+Z(p_3)$.
We define $s$, $t$, and $u$ as in Section 2, and further define 
$s'=(p_1+p_2)^2$, $t'=(p_b-p_2)^2$, and $u'=(p_a-p_2)^2$. 
We choose the color basis $c_1=\delta_{ab} \delta_{12}$ and 
$c_2=T^c_{ba} T^c_{12}$. Then, at one loop, the four elements of the $s$-channel soft anomalous dimension matrix are given by 
\beqa
\Gamma_{S\, 11}^{s\,(1)}&=&C_F \left[\ln\left(\frac{s'-m_t^2}{m_t\sqrt{s}}\right)
-\frac{1}{2}\right] \, ,
\nonumber \\
\Gamma_{S\, 12}^{s\,(1)}&=&\frac{C_F}{2N_c} \ln\left(\frac{t'(t-m_t^2)}{u'(u-m_t^2)}\right) \, ,
\nonumber \\
\Gamma_{S\, 21}^{s\,(1)}&=& \ln\left(\frac{t'(t-m_t^2)}{u'(u-m_t^2)}\right) \, ,
\nonumber \\
\Gamma_{S\, 22}^{s\,(1)}&=&C_F \left[\ln\left(\frac{s'-m_t^2}{m_t \sqrt{s}}\right)-\frac{1}{2}\right]
-\frac{1}{N_c}\ln\left(\frac{t'(t-m_t^2)}{u'(u-m_t^2)}\right)
+\frac{N_c}{2} \ln\left(\frac{t'(t-m_t^2)}{s(s'-m_t^2)}\right) \, ,
\label{Gamma1s}
\eeqa
where $m_t$ is the top-quark mass.

We continue with the $t$-channel processes
$b(p_a)+q(p_b) \rightarrow t(p_1) +q'(p_2)+H(p_3)$ and
$b(p_a)+q(p_b) \rightarrow t(p_1) +q'(p_2)+Z(p_3)$.
We define the kinematical variables as before and  
choose the color basis $c_1=\delta_{a1} \delta_{b2}$ and 
$c_2=T^c_{1a} T^c_{2b}$.
The four elements of the $t$-channel soft anomalous dimension matrix at one loop for these processes are given by 
\beqa
{\Gamma}_{S\, 11}^{t\,(1)}&=&
C_F \left[\ln\left(\frac{t'(t-m_t^2)}{m_t s^{3/2}}\right)-\frac{1}{2}\right] \, ,
\nonumber \\
{\Gamma}_{S\, 12}^{t\,(1)}&=&\frac{C_F}{2N_c} \ln\left(\frac{u'(u-m_t^2)}{s(s'-m_t^2)}\right) \, ,
\nonumber \\
{\Gamma}_{S\, 21}^{t\,(1)}&=& \ln\left(\frac{u'(u-m_t^2)}{s(s'-m_t^2)}\right) \, ,
\nonumber \\
{\Gamma}_{S\, 22}^{t\,(1)}&=& C_F \left[\ln\left(\frac{t'(t-m_t^2)}{m_t s^{3/2}}\right)-\frac{1}{2}\right]
-\frac{1}{N_c}\ln\left(\frac{u'(u-m_t^2)}{s(s'-m_t^2)}\right) 
+\frac{N_c}{2}\ln\left(\frac{u'(u-m_t^2)}{t'(t-m_t^2)}\right) \, .
\label{Gamma1t}
\eeqa

At two loops, the soft anomalous dimension matrices for each of these $t$-channel or $s$-channel processes can be written compactly in terms of the corresponding one-loop results.
We have 
\beqa
\Gamma_{S\, 11}^{(2)}&=& \frac{K}{2} \Gamma_{S\, 11}^{(1)}+\frac{1}{4} C_F C_A (1-\zeta_3) \, ,
\nonumber \\
\Gamma_{S\, 12}^{(2)}&=& \frac{K}{2} \Gamma_{S\, 12}^{(1)} \, ,
\nonumber \\
\Gamma_{S\, 21}^{(2)}&=& \frac{K}{2} \Gamma_{S\, 21}^{(1)} \, ,
\nonumber \\
\Gamma_{S\, 22}^{(2)}&=& \frac{K}{2} \Gamma_{S\, 22}^{(1)}+\frac{1}{4} C_F C_A (1-\zeta_3) \, .
\label{Gamma2s}
\eeqa

We also note that soft anomalous dimension matrices at one loop for processes with three colored particles in the final state have appeared in Refs. \cite{MS,ES}.

To illustrate the usefulness of our formalism, we now briefly apply our methods to the cross section for the $t$-channel process 
$b(p_a)+q(p_b) \rightarrow t(p_1) +q'(p_2)+H(p_3)$. 
NLO calculations for this process have appeared in Refs. \cite{CER,DMMZ}. We set $m_t=173$ GeV and $m_H=125$ GeV, and we use MMHT2014 pdf \cite{MMHT2014} via LHAPDF6 \cite{lhapdf}. The calculations of the cross sections at each perturbative order use the pdf provided at that order. 

In our results below we compute higher-order soft-gluon corrections from resummation at next-to-leading-logarithm (NLL) accuracy, and thus only the terms for the highest two powers of the logarithms are fully determined in our NLO and NNLO expansions. In our discussion, we denote the sum of the LO cross section and the NLO soft-gluon corrections as approximate NLO (aNLO); and we denote the sum of the aNLO cross section and the NNLO soft-gluon corrections as approximate NNLO (aNNLO).

For the $t$-channel $tqH$ production process with scale choice $\mu_F=\mu_R=m_t$, we find aNLO enhancements of the total top + antitop LO cross section of 5.20\% at 8 TeV, 14.9\% at 13 TeV, and 16.2\% at 14 TeV. At aNNLO, we find enhancements over aNLO of 4.3\% at 8 TeV, 4.4\% at 13 TeV, and 4.5\% at 14 TeV. 

The exact NLO enhancements over LO from MadGraph5\_aMC@NLO \cite{MG5} are 5.15\% at 8 TeV, 12.5\% at 13 TeV, and 13.0\% at 14 TeV, which are quite close to the aNLO enhancements, showing that the soft-gluon corrections are a significant and dominant portion of the full corrections, and that our aNLO results approximate very well the exact NLO results at LHC energies. Our results are similarly quite close to those from Refs. \cite{CER,DMMZ} when we use the corresponding pdf sets and parameters used in those references.

A detailed phenomenological study of these processes, including scale dependence, pdf uncertainties, energy dependence, subleading terms, matching to exact NLO, etc., is beyond the scope of this work. We plan to further study these and other processes in future work.

\mysection{Conclusions}

We have presented a soft-gluon resummation formalism for $2 \rightarrow n$ processes in 1PI kinematics, and provided analytical results for the resummed cross section and fixed-orders expansions. We also considered in particular $2 \rightarrow 3$ processes, involving a three-particle final state with a top quark and a Higgs boson, or a top quark and a $Z$ boson, and we provided explicit results for the soft anomalous dimension matrices at one and two loops for those processes, as well as some brief numerical results for $tqH$ production. We foresee a large number of other applications to Standard Model and to Beyond the Standard Model processes. 

\section*{Acknowledgements}
We thank Marco Guzzi for useful discussions on multi-particle kinematics. 
This material is based upon work supported by the National Science Foundation under Grant No. PHY 1820795.

\appendix
\mysection{Frame-invariant integration variables}
\label{sub:invariant_integration_variables}

As shown by Byckling and Kajantie \cite{BKPartKinematics,BKRed}, one can write the expression for the phase space integration of a $2\rightarrow n$ scattering process while integrating over only invariant variables. For processes with massless initial states, we have the phase space integral 
\begin{align}\label{generalPhaseSpace}
R_n(s)=\frac{1}{4s}&\int dp_{1\ldots n-1}^2 dt_{n-1} d\varphi \int dp^2_{1\ldots n-2} dt_{n-2} ds_{n-1} \frac{\Theta(-\Delta_4(n-1))}{8\left[-\Delta_4(n-1)\right]^{1/2}} 
\nonumber \\
& \times \cdots \times  \int dp_{12}^2 dt_{2} ds_{3} \frac{\Theta(-\Delta_4(3))}{8\left[-\Delta_4(3)\right]^{1/2}}
\int dt_1 ds_2 \frac{\Theta(-\Delta_4(2))}{8\left[-\Delta_4(2)\right]^{1/2}}
\end{align}
with $s=(p_a+p_b)^2$ and $p_{1\cdots n}=(p_1+\cdots+p_n)^2$. We define the generalized kinematic invariants $ t_i = (p_a-p_1-\cdots-p_i)^2$, $ u_i = (p_b-p_1-\cdots-p_i)^2$, and $ s_i = (p_i+p_{i+1})^2$. $\Delta_4(i)$ is the four-dimensional Gram determinant which can be written as
\begin{equation}
\Delta_4(i)=\frac{1}{16}\begin{vmatrix}
0 & t_{i-1}-p^2_{1\ldots i-1} & t_i-p^2_{1\ldots i} & t_{i+1}-p^2_{1\ldots i+1} \\
t_{i-1}-p^2_{1\ldots i-1} & 2t_{i-1} & t_i+t_{i-1}-m_i^2 & t_{i-1}+t_{i+1}-s_i \\
t_i-p^2_{1\ldots i} & t_i+t_{i-1}-m_i^2 & 2t_i & t_i+t_{i+1}-m_{i+1}^2 \\
t_{i+1}-p^2_{1\ldots i+1} & t_{i-1}+t_{i+1}-s_{i} & t_i+t_{i+1}-m_{i+1}^2 & 2t_{i+1}
\end{vmatrix}.
\end{equation}
The limits of integration are given by
\begin{align}
\begin{split}\label{Limits}
&p_{1\ldots i}^{2+} = (\sqrt{s}-m_n-\cdots-m_{i+1})^2 \, , \\
&p^{2-}_{1\ldots i} = (m_1+\cdots+m_i)^2 \, , \\
&t_{i-1}'^{\pm} = p^2_{1\ldots i-1}+(2p^2_{1\ldots i})^{-1}\big[(-p^2_{1\ldots i}+t_i)(p^2_{1\ldots i}+p^2_{1\ldots i-1}-m_i^2) \\
&\qquad\ \pm\lambda^{1/2}(p^2_{1\ldots i},t_i,0)\lambda^{1/2}(p^2_{1\ldots i},p^2_{1\ldots i-1},m_i^2)\big] \, , \\
&s_i^\pm = p^2_{1\ldots i-1}+p^2_{1\ldots i+1}+\frac{2}{\lambda(p^2_{1\ldots i},t_i,0)}\bigg[4V(i)\pm\left[G(i)G(i-1)\right]^{1/2}\bigg] \, , 
\end{split}
\end{align}
where $\lambda(x,y,z) = (x-y-z)^2-4yz$, and $G(i)$ and $V(i)$ are given by
\begin{align}
G(i) = -\frac{1}{2}\begin{vmatrix}
0 & 1 & 1 & 1 & 1 \\
1 & 0 & m_{i+1}^2 & t_i & p^2_{1\ldots i} \\
1 & m_{i+1}^2 & 0 & t_{i+1} & p^2_{1\ldots i+1} \\
1 & t_i & t_{i+1} & 0 & 0 \\
1 & p^2_{1\ldots i} & p^2_{1\ldots i+1} & 0 & 0 
\end{vmatrix}
\end{align}
and 
\begin{align}
V(i) = -\frac{1}{8}\begin{vmatrix}
2p^2_{1\ldots i} & p^2_{1\ldots i} - t_i & p^2_{1\ldots i}+p^2_{1\ldots i-1}-m_i^2 \\
p^2_{1\ldots i} - t_i & 0 & p^2_{1\ldots i-1}-t_{i-1} \\
p^2_{1\ldots i+1}+p^2_{1\ldots i}-m_{i+1}^2 & p^2_{1\ldots i+1}-t_{i+1} & 0
\end{vmatrix} \, .
\end{align} 

The angle $\varphi$ describes a rotation of the process around the beam axis and is trivial for our purposes. Integrating it out, including the flux factor and the matrix element $|\mathcal{M}|$, and using the identity $p^2_{1\ldots n-1}=s+t_{n-1}+u_{n-1}-m_n^2$, we obtain the differential partonic cross section
\begin{align}
	s^2\frac{d^2\hat{\sigma}^{ab\rightarrow 1\cdots n}}{dt_{n-1}du_{n-1}} =&\frac{1}{8} \frac{1}{(2\pi)^{3n-5}}\int dp^2_{1\ldots n-2} dt_{n-2} ds_{n-1} \frac{\Theta(-\Delta_4(n-1))}{8\left[-\Delta_4(n-1)\right]^{1/2}} \nonumber \\
& \times \cdots \times  \int dp^2_{12} dt_{2} ds_{3} \frac{\Theta(-\Delta_4(3))}{8\left[-\Delta_4(3)\right]^{1/2}}
\int dt_1 ds_2 \frac{\Theta(-\Delta_4(2))}{8\left[-\Delta_4(2)\right]^{1/2}} |\mathcal{M}|^2 \, .
\end{align}

\mysection{Frame-dependent integration variables}
\label{sub:explicit_angular_dependence}

One can alternatively \cite{BKRed} do the phase space integration of a $2\rightarrow n$ scattering process by breaking the process down into successive $1\rightarrow 2$ decays and integrating over the relevant solid angle in each rest frame explicitly:
\begin{align}\label{generalPhaseSpace2}
R_n(s)=&\int dp^2_{1\ldots n-1} d\Omega_{n} \frac{\lambda^{1/2}(s,p^2_{1\ldots n-1},m_n^2)}{8 s} \nonumber \\
&\times \int dp^2_{1\ldots n-2} d\Omega_{n-1} \frac{\lambda^{1/2}(p^2_{1\ldots n-1},p^2_{1\ldots n-2},m_{n-1}^2)}{8 p^2_{1\ldots n-1}} \nonumber \\
& \times \cdots \times  \int dp^2_{12} d\Omega_{3} \frac{\lambda^{1/2}(p^2_{123},p^2_{12},m_{3}^2)}{8 p^2_{123}}
\int d\Omega_{2} \frac{\lambda^{1/2}(p^2_{12},m_{1}^2,m_{2}^2)}{8 p^2_{12}} \, .
\end{align}

For each of the $1\rightarrow 2$ decays, one takes the center-of-mass (CM) frame of the outgoing particles. For the $n^{th}$ particle, one takes the CM frame of the initial state:
\begin{align}
p_a &= (E_{a}^{(n)},0,0,E_a^{(n)}) \, , \nonumber \\
p_b &= (E_{b}^{(n)},0,0,-E_a^{(n)}) \, , \nonumber \\
p_n &= (E_n,0,|\boldsymbol{p_n}| \sin{\alpha},|\boldsymbol{p_n}| \cos{\alpha}) \, , \nonumber \\
p_{12\ldots n-1} &= (E_{12\ldots n-1},0,-|\boldsymbol{p_n}| \sin{\alpha},-|\boldsymbol{p_n}| \cos{\alpha}) \, .
\end{align}

In this frame, we have
\begin{align}
&E_{a}^{(n)}=E_{b}^{(n)}=\frac{\sqrt{s}}{2} \, , 
& &E_n = -\frac{t_{n-1}+u_{n-1}-2m_1^2}{2\sqrt{s}}=\frac{s+m_1^2-p^2_{1\ldots n-1}}{2\sqrt{s}} \, , \nonumber \\
&|\boldsymbol{p_n}|=\frac{\lambda^{1/2}(s,p^2_{1\ldots n-1},m_n^2)}{2\sqrt{s}} \, , 
& &\cos{\alpha} = \frac{u_{n-1}-t_{n-1}}{\lambda^{1/2}(s,p^2_{1\ldots n-1},m_n^2)}=\frac{s+2u_{n-1}-m_n^2-p^2_{1\ldots n-1}}{\lambda^{1/2}(s,p^2_{1\ldots n-1},m_n^2)} \, ,
\end{align}
and $d\Omega_{n}=d(\cos{\alpha})d\varphi_{n}$, where the integral over $\varphi_{n}$ is trivial for our purposes as before. The first integration can therefore be converted to the frame-independent form
\begin{equation}
\int dp^2_{1\ldots n-1} d\Omega_{n} \frac{\lambda^{1/2}(s,p^2_{1\ldots n-1},m_n^2)}{8 s} = \frac{1}{4 s}\int dt_{n-1} du_{n-1} d\varphi_n 
\end{equation}
to yield the differential partonic cross section
\begin{align}
	s^2\frac{d^2\hat{\sigma}^{ab\rightarrow 1\cdots n}}{dt_{n-1}du_{n-1}} =&\frac{1}{8} \frac{1}{(2\pi)^{3n-5}}\int dp^2_{1\ldots n-2} d\Omega_{n-1} \frac{\lambda^{1/2}(p^2_{1\ldots n-1},p^2_{1\ldots n-2},m_{n-1}^2)}{8 p^2_{1\ldots n-1}} \nonumber \\
& \times \cdots \times  \int dp^2_{12} d\Omega_{3} \frac{\lambda^{1/2}(p^2_{123},p^2_{12},m_{3}^2)}{8 p^2_{123}}
\int d\Omega_{2} \frac{\lambda^{1/2}(p^2_{12},m_{1}^2,m_{2}^2)}{8 p^2_{12}}|\mathcal{M}|^2 \, .
\end{align}

In order to do these integrations, one must go into each $1\rightarrow 2$ frame explicitly. For the $l^{th}$ particle, one takes the frame
\begin{align}
p_a &= (E_{a}^{(l)},0,0,E_a^{(l)}) \, , \nonumber \\
p_b &= (E_{b}^{(l)},0,|\boldsymbol{p_{l+1\ldots n}}|\sin{\psi^{(l)}},|\boldsymbol{p_{l+1\ldots n}}|\cos{\psi^{(l)}}-E_{a}^{(l)}) \, , \nonumber \\
p_{l+1\ldots n} &= (E_{l+1\ldots n},0,|\boldsymbol{p_{l+1\ldots n}}|\sin{\psi^{(l)}},|\boldsymbol{p_{l+1\ldots n}}|\cos{\psi^{(l)}}) \, , \nonumber \\
p_{l} &= (E_{l},0,|\boldsymbol{p_{l}}| \sin{\theta_l}\cos{\varphi_l},|\boldsymbol{p_{l}}| \cos{\theta_l}) \, , \nonumber \\
p_{1\ldots l-1} &= (E_{1\ldots l-1},0,-|\boldsymbol{p_{l}}| \sin{\theta_l}\cos{\varphi_l},-|\boldsymbol{p_{l}}| \cos{\theta_l}) \, , 
\end{align}
with $\int d\Omega_{l} = \int_0^\pi \sin{\theta_l} d\theta_l \int_0^{2\pi} d\varphi_l$. Using the same definitions as above, conservation of momentum and on-mass-shell conditions yield
\begin{align}
&E_{a}^{(l)} = \frac{s-u_{l}-p_{l+1\ldots n}^2}{2 \sqrt{p^2_{1\ldots l}}} \, , & &E_{b}^{(l)} = \frac{s-t_{l}-p_{l+1\ldots n}^2}{2 \sqrt{p_{1\ldots l}^2}}  \, , \nonumber \\
&E_{l+1\ldots n} = \frac{s-p_{1\ldots l}^2-p_{l+1\ldots n}^2}{2 \sqrt{p_{1\ldots l}^2}} \, , & &|\boldsymbol{p_{l+1\ldots n}}| = \frac{\lambda^{1/2}(s,p_{1\ldots l}^2,p_{l+1\ldots n}^2)}{2 \sqrt{p_{1\ldots l}^2}}  \, , \nonumber \\
&E_l = \frac{p_{1\ldots l}^2-p_{1\ldots l-1}^2+p_l^2}{2 \sqrt{p_{1\ldots l}^2}} \, , & &E_{1\ldots l-1} = \frac{p_{1\ldots l}^2+p_{1\ldots l-1}^2-p_l^2}{2 \sqrt{p_{1\ldots l}^2}}  \, , \nonumber \\
&|\boldsymbol{p_l}|=\frac{\lambda^{1/2}(p_{1\ldots l}^2,m_{l}^2,p_{1\ldots l-1}^2)}{2 \sqrt{p_{1\ldots l}^2}} \, , & & \cos{\psi^{(l)}}=\frac{(E_{a}^{(l)})^2-(E_{b}^{(l)})^2+|\boldsymbol{p_{l+1\ldots n}}|^2}{2 |\boldsymbol{p_{l+1\ldots n}}| E_{a}^{(l)}}  \, . 
\end{align}

\end{document}